\documentclass[a4paper]{jpconf}
\usepackage{graphicx}

\usepackage{subfigure}
\usepackage[utf8]{inputenc}
\usepackage[T1]{fontenc}
\usepackage{lmodern}
\usepackage[english]{babel}
\usepackage{float}
\usepackage{microtype}
\usepackage{amsmath}
\usepackage{csquotes}
\usepackage{cite}

\usepackage{lineno}

\def\shat{\ensuremath{\hat{s}}}
\def\that{\ensuremath{\hat{t}}}

\def\sigmahat{\ensuremath{\hat{\sigma}}}

\def\der{\mathrm{d}}

\def\pt#1{\ensuremath{p_{\rm T#1}}}

\def\kt#1{\ensuremath{k_{\rm T#1}}}

\newcommand{\ptq}{\ensuremath{p_{\rm Tq}}}
\newcommand{\pth}{\ensuremath{p_{\rm Th}}}

\newcommand{\alphas} {\ensuremath{\alpha_{s}}}

\newcommand{\xt} {\ensuremath{x_{\rm T}}}

\def\bgi{\begin{itemize}}
\def\endi{\end{itemize}}
\def\bge{\begin{equation}}
\def\ende{\end{equation}}
\def\bgc{\begin{center}}
\def\endc{\end{center}}

\def\kt{{k_{\rm T}}}

\newcommand{\dd}{{\rm d} }
\newcommand{\sqrts}{\sqrt{s}}

\def\pt{p_{\rm T}}

\def\xt{x_{\rm T}}
\def\X{{\rm X}}
\def\neff{n_{_{\rm eff}}}

\def\neff{n}

\def\nnlo{n^{{\rm NLO}}}

\def\sigmamod{\sigma^{\rm model}}

\begin{document}
\title{Studying possible higher twist contributions in the inclusive charged hadron cross sections}

\author{Esko Pohjoisaho}

\address{Helsinki Institute of Physics, P.O. Box 64, FI-00014 University of Helsinki, Finland}
\address{Department of Physics, P.O. Box 35 (YFL), FI-40014 University of Jyväskylä, Finland}

\ead{esko.pohjoisaho@jyu.fi}

\begin{abstract}

In the standard pQCD picture  particles are produced via the parton jet fragmentation process. However, there are also other production mechanisms like higher twist (HT) processes. A usual example of a HT process is a direct production of an outgoing hadron, where the hadron is produced in the hard subprocess without fragmentation.

We study the HT phenomena using a shape analysis ($\xt$ scaling) of the inclusive invariant cross sections of charged hadrons, measured by the ALICE collaboration at center-of-mass energies $\sqrt{s}=$ 2.76 TeV and 7 TeV. The data is compared to PYTHIA8 event generator and to a phenomenological model for HT. Using PYTHIA8, we explore a possible enhancement of HT phenomena for isolated particles, by comparing the shapes of the isolated distributions to inclusive distributions. The results from the standard PYTHIA8, without HT, is compared to a PYTHIA8 where we had included a HT process.

Finally, we found out that the effects observed in the $\xt$ spectra originate from kinematic biases posed by the isolation cuts, rather than from an enrichment of the HT hadrons at the observed cross sections. A more detailed data analysis is ongoing.




\end{abstract}

\section{Introduction}


The cross section for inclusive pion production in pp collisions at $\sqrts = 200$ GeV measured by RHIC \cite{RHICpion} was described well by an NLO pQCD calculation for $\pt > 2$ GeV$/c$. The agreement spanned seven orders of magnitude in the cross section. However, at the LHC energies, deviations of factor two between the NLO and data have been reported by the ALICE collaboration \cite{ALICEppCrossSection} and CMS \cite{Chatrchyan:2011av} (see Fig \ref{xTscaling}). It is already known that selecting a different fragmentation function \cite{Kretzer:2000yf} for the NLO calculation can bring down the deviation with data to 25 \% \cite{d'Enterria:2013vba,Arleo:2010kw}. However, it is not yet fully understood to what extent the so called higher twist (HT) processes, additional to NLO, contribute to the cross section. 
The HT processes have attracted attention also in the field of heavy ion collisions, in the so-called \enquote{Baryon anomaly}, found in measurements of high-$\pt$ baryon production in heavy ion collisions \cite{baryonAnomaly}. So far the question of HT in the LHC is not settled \cite{Rak:1516991}.



\subsection{$\xt$ scaling}
The  kinematics of a $2\rightarrow2$ process is completely fixed by the two dimensional quantities $\sqrts$ and $\pt$, and the two angles $\phi$ and $\theta$. 
By using dimensional analysis and the scaled transverse momentum $\xt=\frac{2\pt}{\sqrts}$, one can express the invariant cross section in a factorized form
\begin{equation}
\label{bbgscaling}
E\ \frac{\dd^3\sigma}{\dd p^3} = \frac{1}{\pt^n} F \left( \frac{2\pt}{\sqrts}, \theta \right)
= \frac{1}{\sqrts^n}G(\xt, \theta),
\end{equation}
where $F$ and $G$ are dimensionless scaling functions. Any combination of $\pt$, $\sqrts$ or the scaled transverse momentum $\xt$ that preserve the dimensions, will result in the scaling form of Eq. \eqref{bbgscaling}. For any scale free theory and vector boson exchange process, the scaling exponent is $n = 4$. However, the experimentally measured $n$ is not a constant $n = 4$, but depends on the studied process, collision energy and the range of $\xt$, reaching values up to $n\sim8$ \cite{arleobrodsky}. For example, the intrinsic partonic transverse momentum $\kt$, QCD radiation, running coupling $\alpha_{s}$, scaling violations in parton distribution functions (PDF) and fragmentation functions (FF) and smearing from the jet fragmentation transverse momentum $j_{T}$ lead into the experimentally observed $n>4$ \cite{arleobrodsky,Tannenbaum:2005by}. For direct photons the exponent is predicted to be roughly one unit smaller due to the absence of fragmentation and one power less in $\alpha_{s}$. This is also supported by the experimental data for direct photons and jets \cite{arleobrodsky}. At the LHC energies the pQCD predicts $n \approx 5\dots6$. 

To explicitly study the behaviour of the scaling exponent, one can choose invariant cross sections from two data sets with different $\sqrts$ at the same values of $\xt$. Using  Eq. \eqref{bbgscaling} for both cross sections allows to cancel out the dimensionless scaling functions to get  
\begin{equation}
n(\xt,\sqrts_1,\sqrts_2) = \frac{\ln(\sigma^{\rm inv}(\xt,\sqrts_2) / \sigma^{\rm inv}(\xt,\sqrts_1) )}{\ln(\sqrts_1 / \sqrts_2)}.
\label{nexpeq}
\end{equation}

\begin{figure}[h!]
\centering
\includegraphics[width=0.5\textwidth] {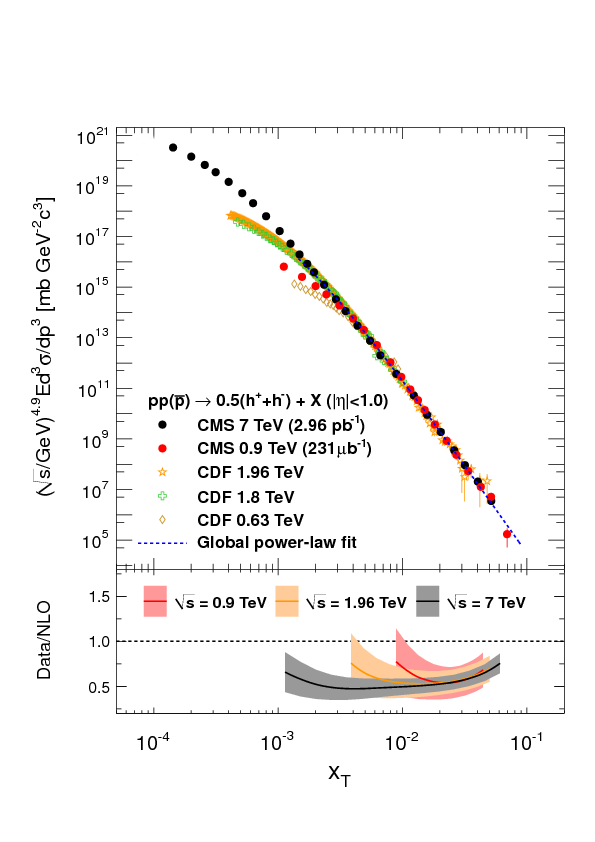}
\caption{(Upper panel) Invariant cross sections as a function of $\xt$ scaled by $(\sqrt{s})^{4.9}$ of inclusive charged particles to demonstrate $\xt$ scaling with different $\sqrts$. (Lower panel) Ratios between data and NLO QCD calculation is shown. Figure from \cite{Chatrchyan:2011av}.}
\label{xTscaling}
\end{figure}

An experimental demonstration of $\xt$ scaling of charged hadron production in proton-proton collisions measured by the CMS collaboration at CERN and CDF collaboration in Fermilab \cite{Chatrchyan:2011av}, along with a global power law-fit, is presented in the upper panel of Fig. \ref{xTscaling}. Cross sections have been scaled by $\sqrt{s}^n$, where the scaling exponent is $n = 4.9$, and one can see curves with different center of mass energies collapse on top of each other in the high $\xt$ range. The black points are for 7 TeV in CMS, red circles for 0.9 TeV in CMS, orange stars for 1.96 TeV in CDF, green crosses for 1.8 TeV CDF and yellow diamonds for 0.63 TeV CDF data. In the lower panel, ratios of data to NLO predictions are shown for different center-of-mass energies \cite{Chatrchyan:2011av}.

\subsection{Higher twist}

The term \enquote{twist} is a historical relic originating from the operator product expansion, which was a tool used to obtain perturbative predictions for deep-inelastic scattering \cite{owens}. Today the leading twist (LT) is understood as standard processes of the pQCD within the collinear factorization, where hadrons are produced via fragmentation processes. In contrast, HT processes are often understood as direct hadron production, where the hadron is produced directly in the hard subprocess without fragmentation \cite{arleobrodsky}. Due to the lack of fragmentation in HT, a production of a direct hadron is argued to involve a large number of active fields, which, according to dimensional counting rules presented in \cite{baryonAnomaly}, lead to steeper $\pt$ spectra and larger scaling exponents $n$ than for the LT hadrons. Therefore the scaling exponents are compared with those obtained from the NLO pQCD \cite{d'Enterria:2013vba}, as was done in \cite{arleobrodsky}. 


The first calculations to estimate the size of the higher-twist contributions were carried out by Bagger and Gunion \cite{Bagger:1981rq}. They found that the pion form factor in the $\gamma q \to \pi q$ process resulted in an extra factor of $1/ \hat{s}$ where $\hat{s}$ is the partonic center-of-mass energy, causing an $n=6$ scaling form for the cross section in Eq. \eqref{bbgscaling}.  The HT contribution has been estimated to be significant especially in the kinematical range of high $\xt$ and low $\pt$, i.e. at low $\sqrts$ \cite{owens,HTcrossSection}. 

Previously, there have been various attempts to experimentally measure the magnitude of the HT effects. Collaborations such as WA77, WA69 and the OMEGA were presenting their findings of direct meson production in a Workshop on High $\pt$ Physics and Higher Twists in 1988. At that time there were many promising results interpreted as evidence of HT, in processes such as $\bar{\nu}N\rightarrow \mu^{+}\pi^{-}X$, Drell-Yann, $\rho$ production and in $e^+ e^-$ data \cite{HigherTwistOMEGA}. 

In search for direct meson production in $\pi^{-}$ Be interactions by the WA77 collaboration \cite{HigherTwistWA77}, the $\rho^0$ production was found to be consistent with LUCIFER Monte Carlo (MC) generator \cite{LUCIFER} with a small HT contribution, whereas the other mesons such as $\phi$ and $K$ were found to be consistent with the purely leading twist MC. Based on MC studies of $\gamma q$ collisions carried out by the WA69 experiment \cite{HigherTwistWA69}, it was found that the shape of the inclusive $\pt$ distribution was not changed significantly by the HT. Based on this remark it would be preferable to take advantage of the expected special kinematical properties of a HT process, by using cuts that enrich the fraction of HT particles in the data. Because the direct meson is created without fragmentation, one could argue that using an isolation cut would enrich the fraction of these mesons in the data, by suppressing the LT mesons originating from fragmentation processes. The effect of isolation was studied in the MC level by the OMEGA Photon Collaboration \cite{HigherTwistOMEGA}, where the meson was not allowed to have any track in its vicinity in a cone of a fixed size. The result was that this kind of isolation did not help significantly in the kinematical range where the HT processes were expected to be significant compared to LT. Still, by including a HT component on top of a QCD based model was in somewhat better agreement with the data than a fit without HT, leaving some room for HT contribution in the interpretation of the data.

More recently, the $\xt$ scaling exponents from world data were studied in \cite{arleobrodsky}. Especially for hadrons at large $\xt$, the exponents were found to be systematically larger than predicted from the NLO pQCD calculations, whereas the photon and jet exponents were in agreement with the theory. The authors suggested that the presence of HT contributions could be a possible explanation for the large exponents as compared to the NLO calculation.

To create predictions for the scaling exponents at RHIC and LHC energies, a two component model was proposed \cite{arleobrodsky} for the cross section
\begin{equation}
\sigma^{\rm model}(p p \to \pi\ \X)\propto \frac{A(\xt)}{\pt^4} + \frac{B(\xt)}{\pt^6},
\label{eq:twocomp}
\end{equation}
where functions $A(\xt)$ and $B(\xt)$ represent LT and HT contributions with their typical $\pt$ dependence. Taking $A$ and $B$ as constants, the effective exponent in this model depends on the relative strength of the HT corrections to the LT cross section, reflected by the ratio $B/A$ \cite{arleobrodsky}. Using Eq. \eqref{eq:twocomp} and the NLO scaling exponent, they constructed an effective exponent
\begin{eqnarray}\label{eq:neff2}
\neff(\xt,\pt,B/A) &\equiv& -\frac{\partial\ln \sigmamod}{\partial\ln\pt}+\nnlo(\xt,\pt)-4\nonumber\\
&=& \frac{2B/A}{\pt^2+B/A}+\nnlo(\xt,\pt).
\end{eqnarray}
In Eq. \eqref{eq:neff2} the terms $\nnlo-4$ have been added to the logarithmic derivative so that the $n$ converges into $\nnlo$ when the HT contributions vanish (i.e. $B\rightarrow0$). This corresponds to the definition of NLO pQCD as a LT in \cite{arleobrodsky}, although from Eq. \eqref{eq:twocomp} the effective exponent for LT contribution would be $n^{\rm model}_{B\rightarrow0} = 4$ for fixed $\xt$, corresponding to a scale free theory. By fitting Eq.~\eqref{eq:neff2} to the world data, the authors of \cite{arleobrodsky} obtained $B/A \sim 50\,\textrm{GeV}^2$. By parametrization of Eq.~\eqref{eq:neff2} they created predictions for the scaling exponents at RHIC and LHC energies. 


\section{Inclusive $\xt$ scaling exponents by ALICE}

Differential cross sections of charged particles in inelastic pp collisions at $\sqrts = 0.9, 2.76$ and 7 TeV have been measured by the ALICE collaboration at the LHC. The measurements were performed in the pseudorapidity range $|\eta| < 0.8$ for particles having $\pt > 0.15$ GeV$/c$ \cite{ALICEppCrossSection}. The data were collected based on tracking information from the Inner Tracking System (ITS) and the Time Projection Chamber (TPC), which are located in the central barrel of the experiment. The minimum-bias interaction trigger was derived using signals from the forward scintillators (VZERO), and the two innermost layers of the ITS, called the Silicon Pixel Detector (SPD) \cite{ALICEppCrossSection}.

\begin{figure} [H]
\subfigure[Scaling exponents $n$]{ 
 \includegraphics[width=0.45\textwidth] {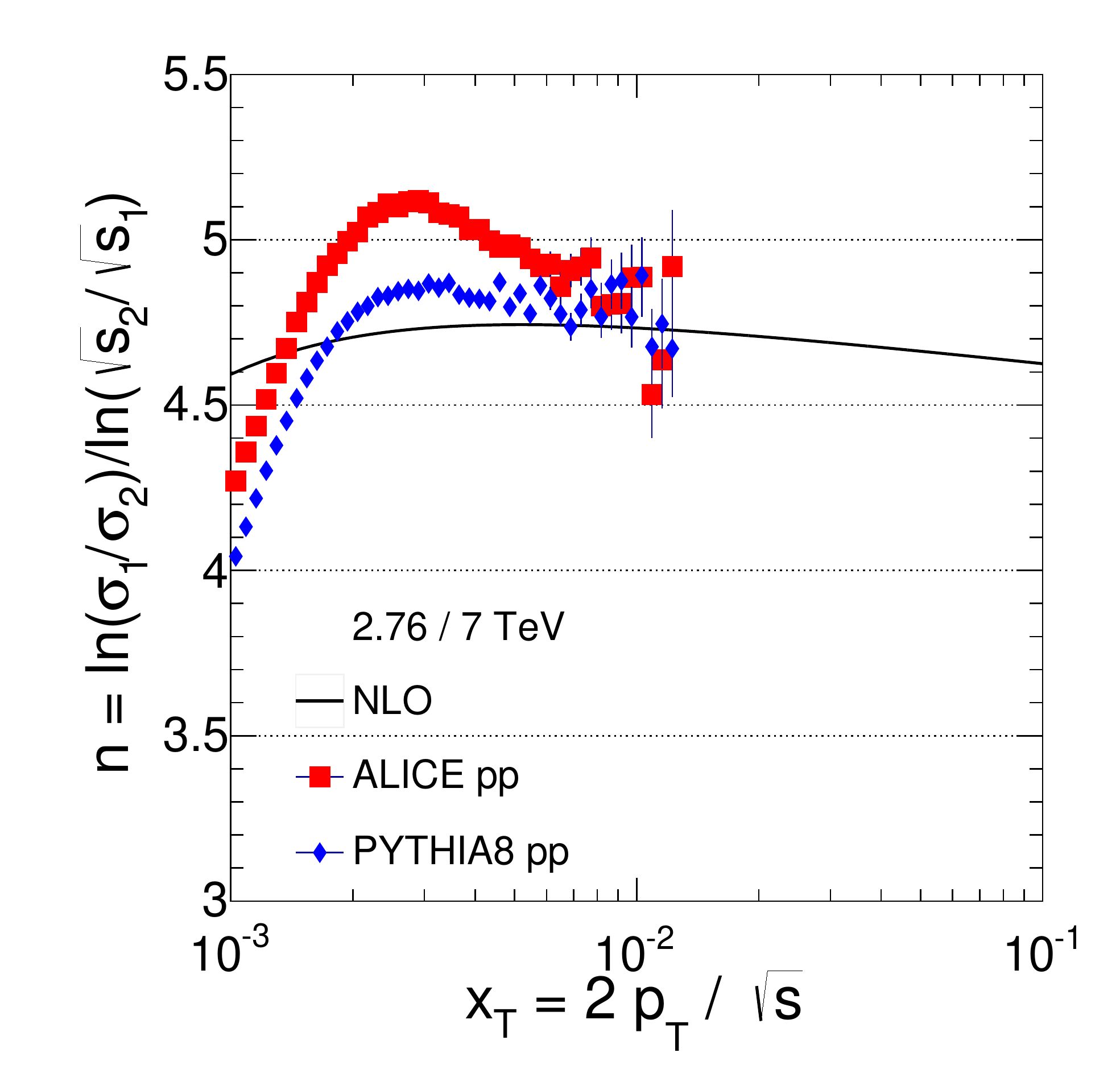} 
} 
\subfigure[Difference to NLO, $\Delta n$]{ 
 \includegraphics[width=0.55\textwidth] {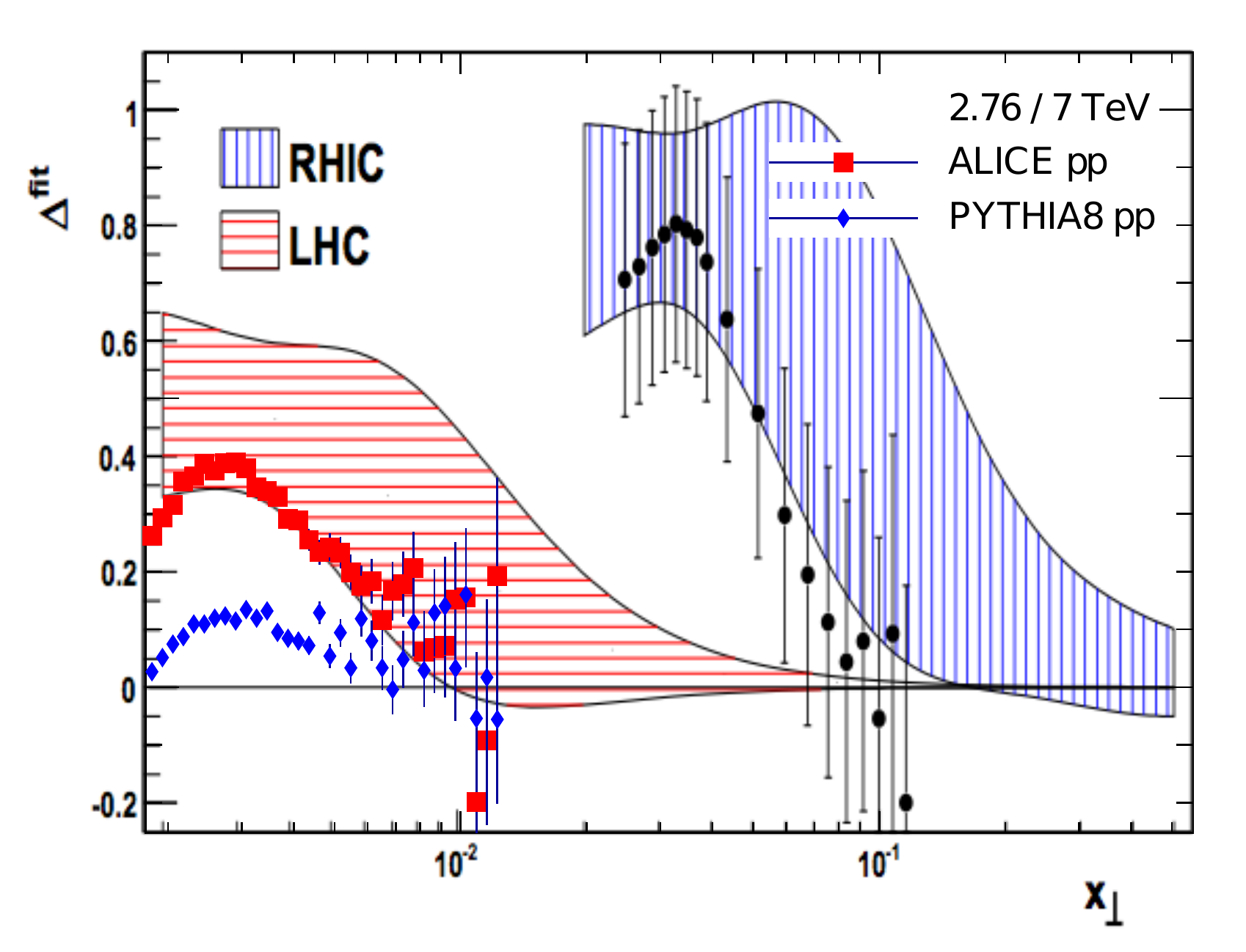} 
} 
\caption{
(a) Scaling exponents $n$ extracted from NLO, ALICE \cite{ALICEppCrossSection} and PYTHIA8 pp data. (b) The difference $\Delta n$ to the NLO exponents (i.e. $\Delta^{\rm fit}$) from ALICE, PHENIX preliminary data \cite{arleoBazilevskyPhenix} and PYTHIA8, with prediction bands based on the two-component model  \cite{arleobrodsky}.
\label{fig:nEff}
}
\end{figure}

In this study, the cross sections as a function of $\xt$ at 2.76 and 7 TeV were derived from the charged hadron $\pt$ spectra measured by ALICE \cite{ALICEppCrossSection}, and the scaling exponents were calculated using Eq.~\eqref{nexpeq} for the $\xt$ spectra. In Fig. \ref{fig:nEff} (a), the scaling exponents are plotted as a function of $\xt$ for the NLO calculation \cite{d'Enterria:2013vba} (black line), default PYTHIA8 simulation \cite{Sjostrand:2007gs} (blue points), and for the exponents calculated from the inclusive charged particle data measured by the ALICE collaboration \cite{ALICEppCrossSection} (red points) at 7 TeV and 2.76 TeV.

In Fig. \ref{fig:nEff} (b) the NLO exponents have been subtracted from the ALICE data and PYTHIA8 exponents from Fig. \ref{fig:nEff} (a), to compare the results with prediction bands from \cite{arleobrodsky}, where the prediction was based on further parametrization of Eq.~\eqref{eq:twocomp} for the difference to the NLO exponent, i.e. $\Delta^{\mathrm{fit}} = n_{\mathrm{measured}} - n_{\mathrm{NLO}}$. The red band is the prediction for the LHC energies and the blue band for RHIC energies \cite{arleoNew}. The black points are preliminary data from the PHENIX collaboration \cite{arleoBazilevskyPhenix}, blue points from the default PYTHIA8 and red points the ALICE data. 


The data points in Fig. \ref{fig:nEff} (b) seem to be in agreement with the prediction. Preliminary data points from PHENIX have been shown to be in agreement with the prediction band for RHIC energies \cite{arleoNew}, but until now the inclusive ALICE data points have not been shown on top of the prediction for LHC energies. The points from PYTHIA8 are closer to the NLO than the ALICE data, but still non-zero. In fact, it was seen that PYTHIA8 describes the measured shapes of the cross sections well but the absolute normalization disagrees with the ALICE data ~$20-25$ \% (not shown here). If the $\pt$ cross sections were manually rescaled closer to data, the resulting scaling exponents $n$ would (naturally) agree with the measured data, i.e. the difference between the data and PYTHIA8 on Fig. \ref{fig:nEff} (b) originates from the fact that this PYTHIA tune misses the absolute normalization of the cross sections.



\section{PYTHIA implementation of higher twist}


There are no HT processes, like direct hadron production, implemented in the standard PYTHIA8 \cite{Sjostrand:2007gs}. A process of this kind was added \cite{TorbjornPrivate} into PYTHIA8 using the elementary cross section for a HT direct pion production process $(q_A g \rightarrow q_B \pi)$ computed in \cite{HTcrossSection}. The cross section is
\begin{equation}
\frac{\der \sigmahat}{\der \that} = \frac{\pi \alphas^2}{\shat^2} \frac{\alphas}{4\pi} \frac{s_0}{\shat} F(\cos \theta),
\label{HTxsect}
\end{equation}
where $s_0 = 16 \pi^2 f_{\pi}^2$ is a higher twist scale, fixed by the pion weak decay constant $f_{\pi}$, and $F(\cos \theta)$ is the angular function for the c.m. scattering angle $\theta$, $\cos \theta = 1+2\that/\shat$, and has a form 
\begin{equation}
F(z) = \frac{2}{27}(1-z) \left( 1 + \frac{4}{(1+z)^2} \right).
\end{equation}

The effects of using an isolation cut was studied with PYTHIA8. The isolation condition was checked for every particle by adding up the $\pt$ of other final charged particles around it inside a cone of radius $R=\sqrt{\Delta \eta^2 + \Delta \phi^2}=0.4$. To declare a particle as isolated, the $\pt$ sum in the cone was required to be less than 10\,\% of the particle's $\pt$. It has been suggested that isolation cut would enrich the fraction of HT in the data. As a result, the scaling exponents $n$ obtained from the isolated $\xt$ spectra were anticipated to be larger than in the inclusive case \cite{arleobrodsky}.

\begin{figure} [H]
\centering
 \includegraphics[width=0.45\textwidth] {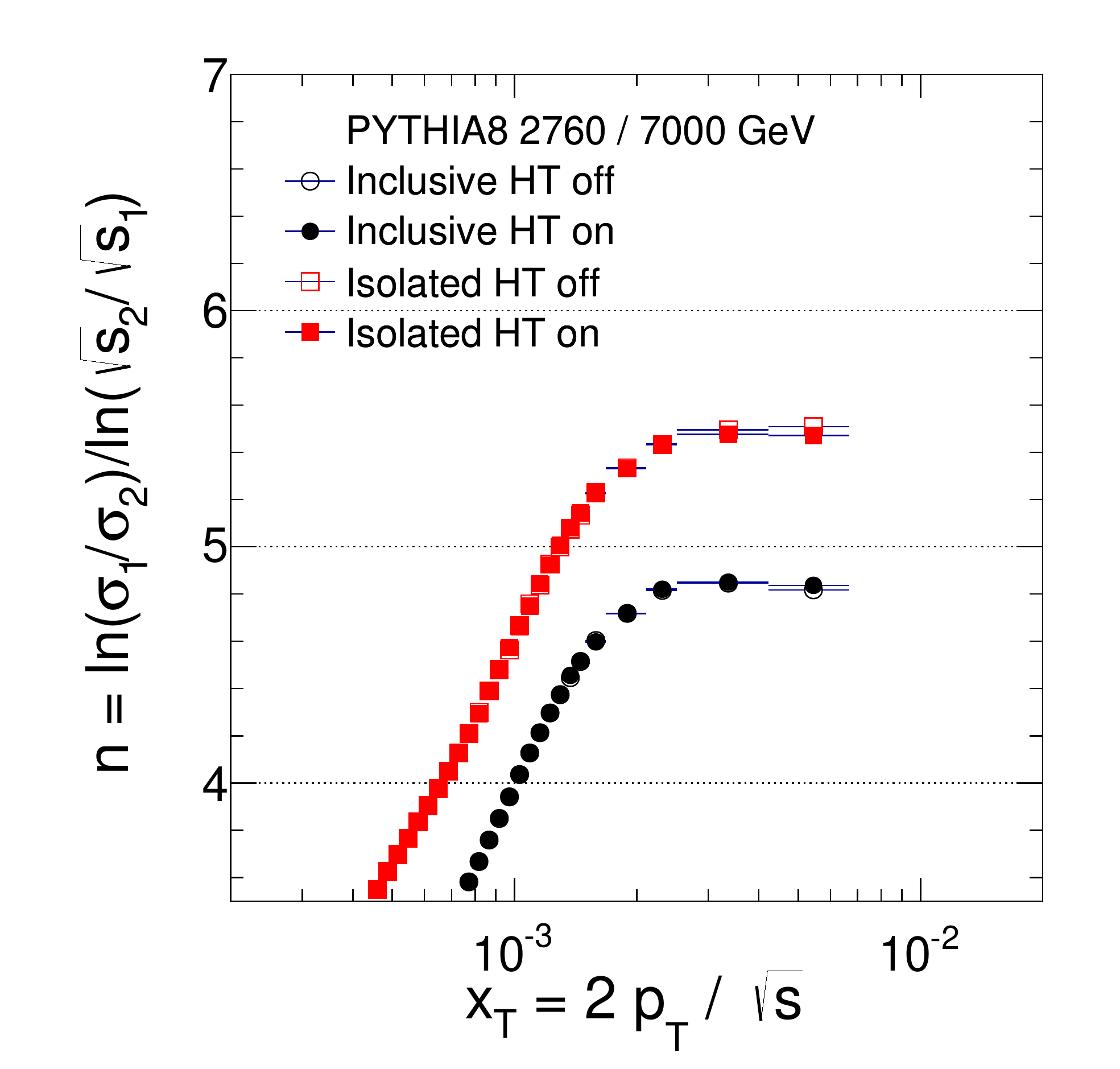} 
\label{fig:PYTHIAHT}
\caption{
Scaling exponents from PYTHIA8 with Higher twist process on/off for inclusive and isolated hadrons.
}
\end{figure} 


In Fig. \ref{fig:PYTHIAHT} one can observe that the effect of the HT process depicted by Eq.~\eqref{HTxsect} to the exponent $n$ in PYTHIA8 is quite negligible. In fact, there was on the average only one HT pion per 10000 charged hadrons in the simulation and applying the isolation cuts enriched the ratio only by a factor of 2. This suggests that the HT process described in \cite{HTcrossSection} does not have a significant role in LHC energies.

One can see from Fig. \ref{fig:PYTHIAHT} that the isolated spectrum has larger $n$ than the inclusive spectrum, even without any HT processes, suggesting a kinematical bias caused by the isolation cut itself. This effect was further studied by a toy Monte Carlo model based on the \enquote{the parent-child relationship} (PCR), presented by Bjorken in  \cite{Bjorken:1973kd}. Assuming that the fragmenting partons follow a power law distribution, $\der N/\der \ptq \sim \ptq^{-(p-1)}$, where the $\ptq$ is the parton $\pt$, the final hadron spectrum follows asymptotically the same shape
\begin{equation}
\frac{1}{\pth}\frac{\der N}{\der \pth} = \frac{1}{\pth^p} \int_{\xt}^{1} z^{p-2} D(z) \der z,
\label{ToyMCdNdpt}
\end{equation}
where $\pth$ is the transverse momentum of the hadron and $D(z)$ is the fragmentation function. In the toy MC model we use the power law exponent $p=6$  for the \enquote{2.76 TeV} and $p=5$ for the \enquote{7 TeV} distributions, that roughly give the asymptotic behaviour of the measured data \cite{ALICEppCrossSection}. For the fragmentation function, we took for simplicity $D(z) \sim \exp(-8.2 z)$, as was done in \cite{Adler}.

\begin{figure} [H]
\centering
\subfigure[Isolation fractions]{ 
\includegraphics[width=0.45\textwidth] {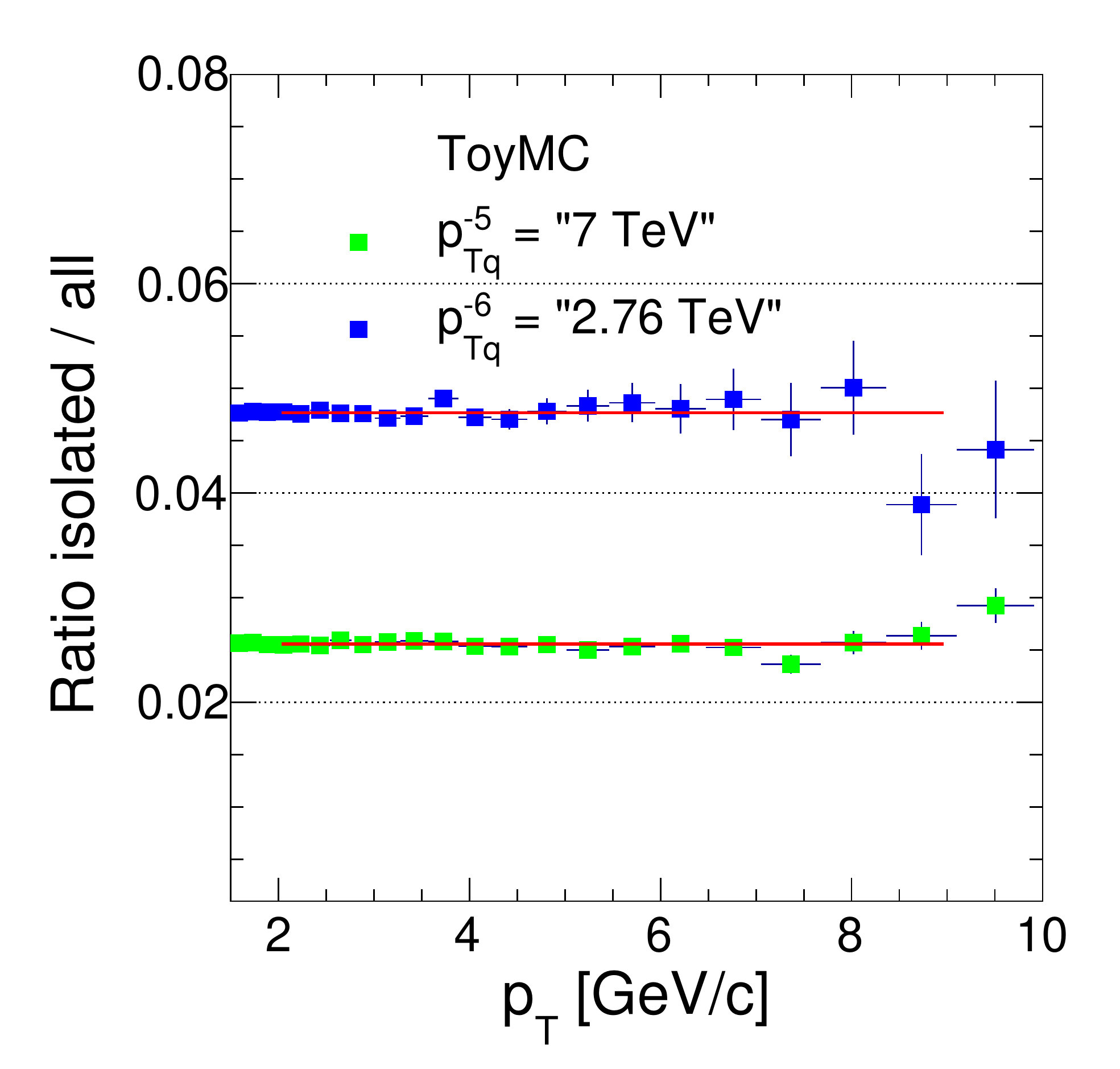} 
}
\subfigure[Increase of the scaling exponent $n$]{ 
\includegraphics[width=0.45\textwidth] {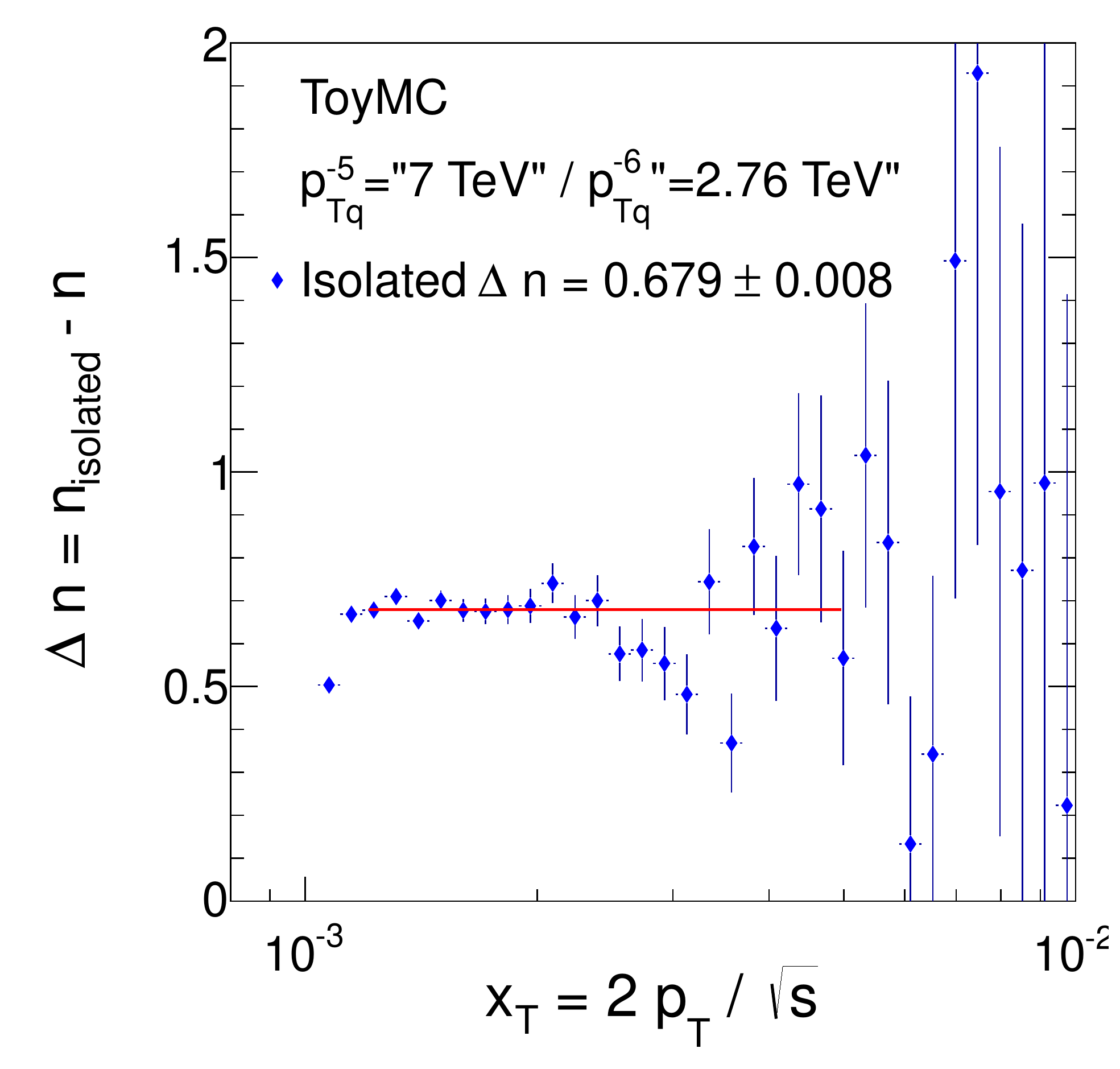} 
}
\caption{(a) Isolation fractions obtained from the toy MC model for a parton cascade. (b) The difference $\Delta n$ caused by the isolation cut in the toy MC.}
\label{fig:ToyMCisolProb}
\end{figure}

To get a simple estimate of the isolation effects corresponding 10 \% cone activity, we first force $z > 0.9$ in the integral in Eq.~\eqref{ToyMCdNdpt}, which yields a constant isolation fraction
\begin{equation}
\frac{\rm isolated}{\rm inclusive} = \frac{\int_{0.9}^{1} z^{p-2} D(z) \der z}{\int_{\xt\approx0}^{1} z^{p-2} D(z)  \der z}.
\label{PCRisorProb}
\end{equation}
By inserting fractions obtained from Eq.~\eqref{PCRisorProb} into Eq.~\eqref{nexpeq}, we found out that the scaling exponent is increased by 0.77 as compared to the non-isolated case. In addition, a cascade process was used. By repeatedly sampling a fragmentation variable $z$ from the exponential \enquote{fragmentation function}, one could create a collection of hadrons from the initial parton. The isolation condition for the leading hadron could then be checked by comparing the hadron $\pt$ to the $\pt$ sum of all the other hadrons in the toy MC event. It was observed that mostly the high $z$ hadrons survive the isolation cut. Isolation probabilites obtained with this method are shown in Fig. \ref{fig:ToyMCisolProb} (a), and the increase of the scaling exponents,  $\Delta n = n_{\rm isolated} - n$, is shown in Fig. \ref{fig:ToyMCisolProb} (b). Using the isolation fractions from Fig. \ref{fig:ToyMCisolProb} (a) with Eq.~\eqref{nexpeq}, the scaling exponent increased by $0.679 \pm 0.008$, marked by the red line in Fig. \ref{fig:ToyMCisolProb} (b). This result is of similar magnitude as with the integral method. More importantly, the increase of the exponent in the toy MC model is very close to the increase we observed in the standard PYTHIA8, see Fig. \ref{fig:PYTHIAHT}.


\section{Conclusion}

The $\xt$ scaling exponents $n$ for inclusive charged hadrons were extracted from ALICE pp data, PYTHIA8 and NLO at $\sqrts = 7$ TeV and 2.76 TeV. The scaling exponents obtained from the ALICE data are in agreement with the two-component model predictions. It was shown that PYTHIA8 describes the shape of the $\pt$ spectra quite well, but the normalization is off by $\sim20\,\%$. If this normalization were correct, the scaling exponents would be similar to data. As it was expected that the HT contribution would steepen the spectra and thus increase $n$, the similar shape of the data with PYTHIA8 may suggest that the HT contribution is not very large.


An MC level study of HT contribution was made by implementing a direct $\pi^{\pm}$ production process \cite{HTcrossSection} into PYTHIA8 \cite{TorbjornPrivate}. It turned out that the fraction of HT pions of all charged hadrons was roughly 1/10000 in the simulation, and applying the isolation cuts did not significantly enrich the ratio. Hence, the direct pion production as described by \cite{HTcrossSection} gives a negligible contribution to the cross section and leads into barely visible effects in both inclusive and isolated scaling exponents.


The isolation cut was seen to increase $n$ due to a kinematical bias in two ways. Firstly, by biasing the fragmentation into large values of $z$ and thus making the spectra steeper. Secondly, the $\pt$ cross section is harder at 7 TeV than at 2.76 TeV, which lead into higher probability for a hadron to be isolated at 2.76 TeV collision energy. The difference in the isolation probabilities directly results in an increase of the scaling exponent. It seems that the observed effect could be largely explained by these kinematical biases. A more careful study would be needed to verify to what extent there is room for HT contributions in the data. Also, the isolation study is so far done only with PYTHIA8, while the analysis for the ALICE data is ongoing.


\ack

We are most grateful to Torbjörn Sjöstrand for his kind help in implementing the direct hadron production process into PYTHIA8. Special thanks to Hannu Paukkunen from the University of Jyväskylä for providing us the NLO pQCD cross section spectra used in this study. Finally, thanks to the Helsinki Institute of Physics for funding the research and making it possible to participate in the 9th International Workshop on High-pT Physics.


\section*{References}
\providecommand{\newblock}{}

\end{document}